\documentclass[sigconf]{acmart}

\usepackage{fancyhdr}
\usepackage{graphicx}
\usepackage{epstopdf}
\usepackage{multirow}
\usepackage{float}

\floatstyle{plaintop}
\restylefloat{table}

\AtBeginDocument{%
  \providecommand\BibTeX{{%
    \normalfont B\kern-0.5em{\scshape i\kern-0.25em b}\kern-0.8em\TeX}}}

\copyrightyear{2021} 
\acmYear{2021} 
\setcopyright{acmcopyright}\acmConference[KDD '21]{Proceedings of the 27th ACM SIGKDD Conference on Knowledge Discovery and Data Mining}{August 14--18, 2021}{Virtual Event, Singapore}
\acmBooktitle{Proceedings of the 27th ACM SIGKDD Conference on Knowledge Discovery and Data Mining (KDD '21), August 14--18, 2021, Virtual Event, Singapore}
\acmPrice{15.00}
\acmDOI{10.1145/3447548.3467191}
\acmISBN{978-1-4503-8332-5/21/08}

\begin{document}
\fancyhead{}

\title{DMBGN: Deep Multi-Behavior Graph Networks for Voucher Redemption Rate Prediction}

\author{Fengtong Xiao}
\email{fengtong.xiao@lazada.com}
\affiliation{%
  \institution{Alibaba Group}
  \country{Singapore}
}
\author{Lin Li}
\authornote{Both authors contributed equally to this work.}
\email{boolean.ll@alibaba-inc.com}
\author{Weinan Xu}
\authornotemark[1]
\email{stella.xu@lazada.com}
\affiliation{%
  \institution{Alibaba Group, Singapore}
  \country{}
}
\author{Jingyu Zhao}
\email{jingyu.zhao@lazada.com}
\affiliation{%
  \institution{Alibaba Group}
  \country{Singapore}
}
\author{Xiaofeng Yang}
\email{xiaofeng.yang@lazada.com}
\affiliation{%
  \institution{Alibaba Group, Singapore}
  \country{}
}
\author{Jun Lang}
\email{bill.lang@lazada.com}
\affiliation{%
  \institution{Alibaba Group, Singapore}
  \country{}
}
\author{Hao Wang}
\email{longran.wh@alibaba-inc.com}
\affiliation{%
  \institution{Alibaba Group, Singapore}
  \country{}
}

\begin{abstract}
In E-commerce, vouchers are important marketing tools to enhance users' engagement and boost sales and revenue. The likelihood that a user redeems a voucher is a key factor in voucher distribution decision. User-item Click-Through-Rate (CTR) models are often applied to predict the user-voucher redemption rate. However, the voucher scenario involves more complicated relations among users, items and vouchers. The users' historical behavior in a voucher collection activity reflects users' voucher usage patterns, which is nevertheless overlooked by the CTR-based solutions. In this paper, we propose a Deep Multi-behavior Graph Networks (DMBGN) to shed light on this field for the voucher redemption rate prediction. The complex structural user-voucher-item relationships are captured by a User-Behavior Voucher Graph (UVG). User behavior happening both before and after voucher collection is taken into consideration, and a high-level representation is extracted by Higher-order Graph Neural Networks. On top of a sequence of UVGs, an attention network is built which can help to learn users' long-term voucher redemption preference. Extensive experiments on three large-scale production datasets demonstrate the proposed DMBGN model is effective, with 10\% to 16\% relative AUC improvement over Deep Neural Networks (DNN), and 2\% to 4\% AUC improvement over Deep Interest Network (DIN). Source code and a sample dataset are made publicly available to facilitate future research\footnote{https://github.com/fengtong-xiao/DMBGN}.
\end{abstract}



\maketitle
\vspace{-0.2cm}

\section{Introduction}
In E-commerce, vouchers have already become significant tools which not only drive sales, but also help user growth and enhances customer loyalty. An online voucher (or coupon in some literature) is a digital ticket that can be redeemed with discount or rebate when a product is purchased\footnote{Retailers may limit the scope of a voucher to a specific pool of promoted items. This paper focuses on those voucher applicable to all the products in the platform.}. A typical voucher is characterized by two components: condition and discount. The former specifies requirements of a voucher to be applicable, e.g, the minimum number of items purchased or the minimum amount of money spent in an order, and the latter refers to the amount or percentage deductible when the voucher is applied. This work focuses on minimum-spend and discount-amount based vouchers. 

\begin{figure}[H]
  \centering
  \hspace*{-0.2cm} 
  \vspace*{-0.1cm}
  \includegraphics [width=3.5in]{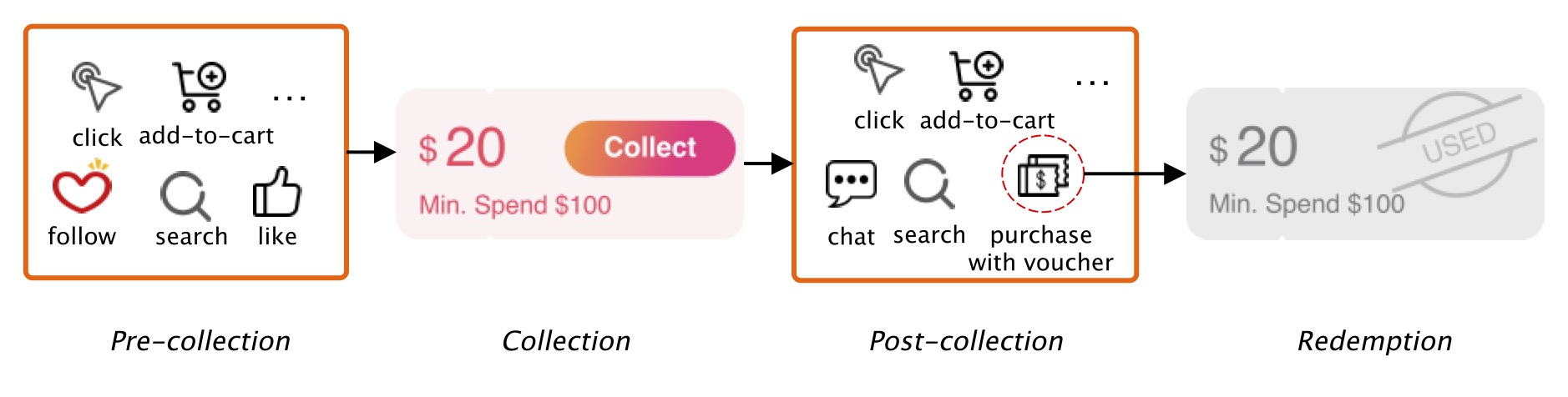}
  \setlength{\abovecaptionskip}{-0.2cm}
  \setlength{\belowcaptionskip}{-0.3cm}
  \caption{Illustration of interactions between users and vouchers in different phases.}
  \label{fig:lifecycle_all}
\end{figure}

Fig. \ref{fig:lifecycle_all} shows a typical user-voucher interaction scenario on an E-commerce platform: a user may do some online shopping before being assigned a voucher (i.e., pre-collection phase). After collecting the voucher (either manually or automatically), the user may continue to search, click or add-to-cart the items that he/she wants to buy with the voucher (i.e., post-collection phase). Finally, the user may redeem the voucher by purchasing some items (i.e., redemption phase), or simply leave the voucher expired without redemption.

\vspace*{-0.3cm}

The capability of predicting the user-voucher redemption rate (VRR henceforth in this paper), the likelihood that a user may redeem a voucher, would contribute to a successful voucher marketing in multiple ways:

\begin{itemize}
    \item \textbf{\textit{Performance Forecasting}}: Retailers are able to estimate the returns of a voucher (e.g. overall redemption rate, redeemed voucher count) before a campaign starts. 
    \item \textbf{\textit{Budget Control}}: Retailers are able to know when to stop distributing vouchers so that the marketing budget would not be over-utilized or under-utilized.
    \item \textbf{\textit{Personalized Distribution}}: Retailers are able to distribute different voucher schema (e.g. condition and discount) to different customers to maximize the overall returns, by applying an optimization algorithm to the VRR results.
\end{itemize}

There are some preliminary studies \cite{hu2018promotion, li2020spending, qiongyu2020prediction} focusing on VRR prediction. Those studies solve voucher redemption rate prediction tasks by directly borrowing the ideas from user-item Click-Through-Rate (CTR) prediction models \cite{zhou2018deep, cheng2016wide}. A common practice is to treat a given voucher in the same way as treating an item, simply replacing item features by voucher features in model training and prediction.

Nonetheless, the VRR task is more challenging in several ways. Firstly, in item-wise CTR models, items in historical behavior and the item to be evaluated are homogeneous, while in VRR, the behavior involves heterogeneous relations between vouchers and items. Secondly, in CTR models, users' behavior on items has no obvious inter-dependent relationships. In voucher scenarios, by contrast, the user behavior is often triggered by the collected voucher and is strongly affected by the voucher condition and discount. As shown in Table \ref{tab:user_seq}, the behavior sequence in the post-collection phase becomes relatively longer than in pre-collection. Lastly, a user may have a multiple voucher redemption history, the long-term and short-term voucher usage preferences of the user need to be considered and balanced to better predict the VRR of the current voucher to be distributed (namely target voucher). 

To tackle the above issues and shed light on the VRR prediction task, we propose a graph-based model Deep Multi-Behavior Graph Networks (DMBGN). To evaluate a user’s VRR of the target voucher, DMBGN takes into consideration the user’s historical interactions with vouchers. To model the structural user-voucher-item relationships, a User-Behavior Voucher Graph (UVG) is constructed for each user-voucher interaction activity. A UVG is a satellite-like heterogeneous graph containing nodes for both voucher and items in user behavior happening before and after the voucher collection. The techniques of Higher-order Graph Neural Networks are adopted to extract the high-level structural representation of a voucher activity, so that relationships like user-item preference, voucher-item dependency and user-voucher preference can be better distilled. To learn the user's long-term interests in voucher redemption, an attention network is applied on top of the historical UVG sequence. Finally, all the implicit representations of user, item, voucher and other side information are forwarded to a DNN network for the VRR prediction.

The main contributions of this paper are as follows:

\begin{itemize}
    \item To our best knowledge, this is the first work to propose a model specialized for the VRR prediction task. The model, DMBGN, follows the user's mindset in a real voucher activity and represents the historical behavior happening both before and after voucher collection with a User-Behavior Voucher Graph (UVG).
    \item DMBGN is also the first work to utilize techniques of graph neural networks to learn the complex user-voucher-item relationships, and exploits an attention network to mine users' long-term voucher redemption preference with respect to the target voucher.
    \item Extensive experimental results conducted on three large-scale production datasets collected from E-commerce platform Lazada indicate promising performance of DMBGN. Compared with Deep Neural Networks (DNN), DMBGN achieves 10\% to 16\% relative AUC improvement. Compared with Deep Interest Network (DIN), DMBGN achieves 2\% to 4\% uplift in AUC as well.
\end{itemize}
\vspace{-0.2cm}

\begin{table}
  \caption{The average sequence lengths and the percentage difference of items from user behavior (click, add-to-cart and order) happening before and after voucher collection.}
  \label{tab:seq length}
  \footnotesize
  \begin{tabular}{cccc}
    \toprule
    \multirow{2}{*}{User Behavior} & Avg. Sequence & Avg. Sequence & Avg. Sequence Length\\
    & Length (before) & Length (after) & Diff \% = (after)/(before)-1\\
    \midrule
    Clicked Items & 79.8 & 92.2 & 15.5$\%$\\
    Add-to-cart Items & 10.9 & 13.7 &25.2$\%$\\
    Ordered Items & 1.8 & 3.6 &105.7$\%$\\
  \bottomrule
\end{tabular}
\label{tab:user_seq}
\vspace{-0.25cm}
\end{table}

\section{Related Work}
\subsection{Voucher-related Works}
There are some preliminary studies in the literature addressing voucher-related problems. Those works can be divided into two types: i) the VRR prediction and ii) voucher distribution decision-making. For the first type, most of the studies adopt the CTR models widely utilized in item-wise recommendation. In \cite{qiongyu2020prediction}, XGBoost model is used to solve the Online-to-Offline coupon redemption rate prediction, based on both user profile and coupon characteristic features. In \cite{li2020spending}, a sequence-based structure with attention-based mechanism is deployed, which considers long-term and short-term user behavior in addition to the voucher information. As discussed in introduction, our proposed model can better capture the more complex user-voucher-item relations with graph techniques, which is not presented in the previous works. For the second type, the task is to assign appropriate vouchers to different users based on the VRR prediction results and the total budget constraints. The most conventional optimization method for this task is the Multiple-Choice Knapsack Problem (MCKP) algorithm \cite{kellerer2004multiple}. Some online learning algorithms can also be applied, such as online-MCKP \cite{chakrabarty2008online}, linear contextual bandits with knapsacks \cite{agrawal2015linear} and reinforcement learning based algorithms \cite{gai2018optimal}. In this work, we focus on the first type of the problem on VRR prediction.

\subsection{CTR and CVR Prediction Models}
User-item Click-Through-Rate (CTR) prediction, which is to predict the probability of a user clicking an item, is the key task in the item-wise recommendation domain. The widely utilized models include Wide-and-Deep (WDL) \cite{cheng2016wide} and Deep Factorization Machine (DFM) \cite{DeepFM}. In recent years, attention-based structures to capture user's historical interest representation become a hot topic. Among them, Deep Interest Network (DIN) \cite{zhou2018deep}, Deep Evolution Interest Network (DIEN) \cite{zhou2019deep}, Deep Hierarchical Attention Network (DHAN) \cite{xu2020deep}, Deep Cross Networks (DCN) \cite{wang2017deep} and Behavior-sequence Transformer (BST) \cite{chen2019behavior} are the representatives. There, the target and the attended objects are all items, and the behavior usually has no strong dependent relations. However, the VRR task needs to consider heterogeneous and inter-dependent relationships among users, items and vouchers.

Conversion rate (CVR), the possibility that a user will finally convert (e.g. purchase the item), is another important task in item-wise recommendation. The multi-task learning framework is a popular solution in this area. In \cite{ma2018modeling, bao2020gmcm}, MMoE and GMCM models utilize the shared bottom structure and task-specific upper structures to optimize CTR and CVR together. In \cite{wen2020entire}, ESMM2 follows a similar strategy but instead optimizes CTR and Click-Through-Conversion Rate (CTCVR). Although multi-task learning can help to estimate the conversion before the target is exposed, it is not appropriate for the VRR task. Users usually have no objection to collect an assigned voucher, and under some circumstances, the voucher is auto-collected according to the platform's strategies. Therefore, it is difficult or even impossible to co-train the collection rate and the redemption rate in the multi-task manner. 

There are also some works which take multiple user behaviors into account for more accurate CTR and CVR prediction. In these works, various behaviors such as clicks, favorites, add-to-cart, purchases, etc are considered and specific techniques are applied to model user-item relationships for better inference, such as hetergeneous graph convolution network \cite{bao2020gmcm, jin2020multi}, cascading ordinal behavior structure \cite{Chen2020EfficientHC, 8731537} and transformer architecture \cite{xia2020multiplex}. However, our work for VRR aims to model user-voucher-item relationships through a graph network, which focuses more on capturing different user behavior patterns happening before and after voucher collection.
\vspace{-0.05cm}

\subsection{Graph Embedding and Related Models}
Recent years have seen the surge of research interests in deep graph-network techniques. A series of graph embedding techniques such as DeepWalk \cite{perozzi2014deepwalk}, Node2Vec\cite{grover2016node2vec} and MetaPath2Vec \cite{dong2017metapath2vec} have been widely utilized in various applications. More recently, convolutional network structures become popular which can dig deeper structural relationships from the graph. A number of models such as WL GNN \cite{morris2019weisfeiler}, GraphSage \cite{hamilton2017inductive}, Graph-Attention Networks (GAT) \cite{velivckovic2017graph}, Graph Convolutional Networks (GCN) \cite{kipf2016semi} and Gated Graph Neural Networks (GGNN) \cite{li2015gated} have been successfully deployed in the fields of social networking, advertising, risk assessment and recommendation systems. For the CTR tasks, DHGAT\cite{niu2020dual} employs a heterogeneous graph structure to model the relationships among users, queries, shops and items. ATBRG \cite{feng2020atbrg} utilizes graph pruning techniques to get more effective recommendations. GMCM \cite{bao2020gmcm} adopts GCN to generate the hidden representations of various types of behavior such as click, review and add-to-cart.
\vspace{-0.05cm}

\section{Methodology}
\vspace{-0.1cm}
\begin{figure}[htb]
  \centering
  \includegraphics[width=3.4in]{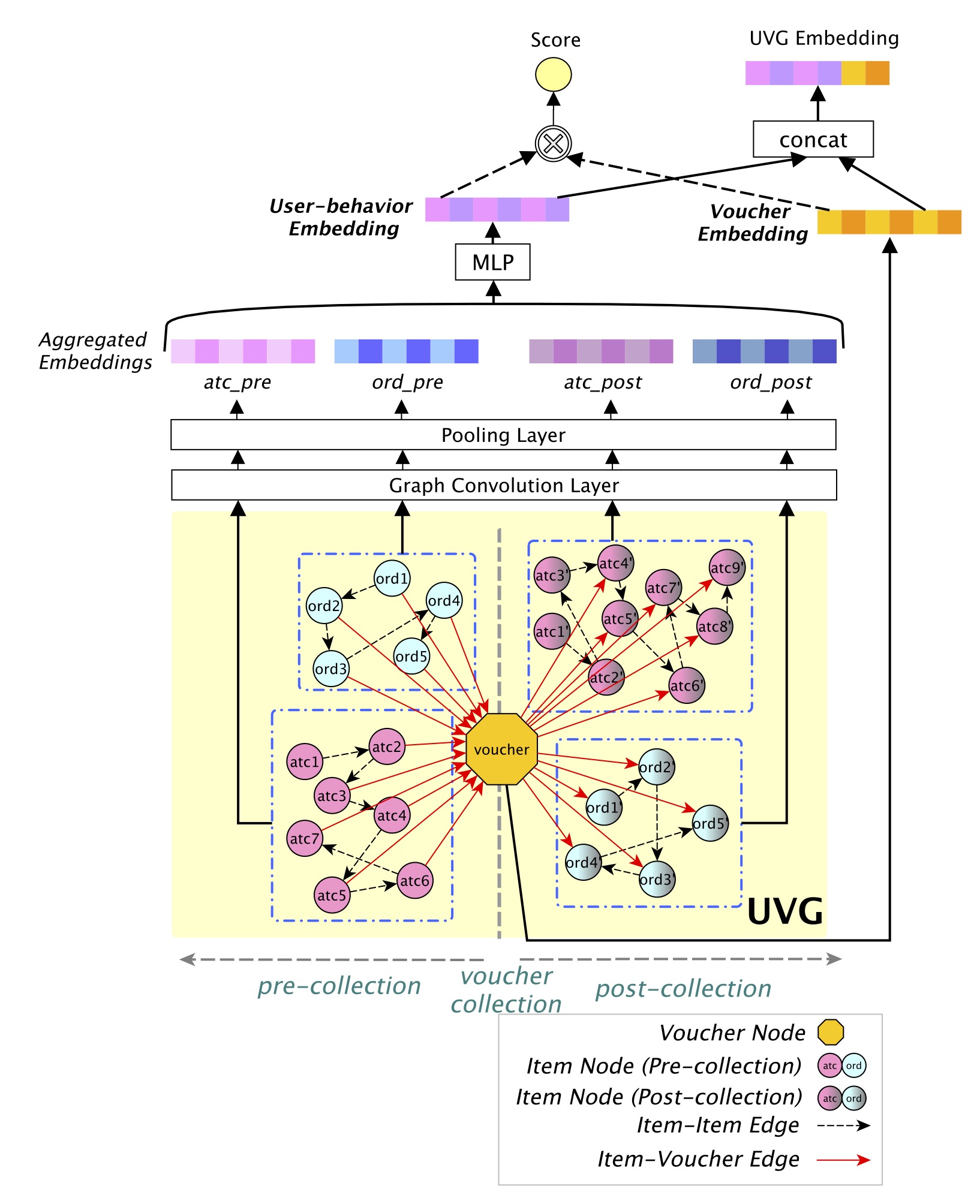}
  \setlength{\abovecaptionskip}{-0.5cm}
  \setlength{\belowcaptionskip}{-0.5cm}
  \caption{Illustration of User-Behavior Voucher Graph (UVG). The voucher node in the center represents the collected voucher. Item nodes from user behavior are sub-grouped by action types ($atc$ and $ord$) and phases (pre-collection and post-collection).}
  \label{fig:sub-graph}
  \vspace{-0.4cm}
\end{figure}

\begin{figure*}[h]
  \centering
   \vspace*{-0.4cm} 
   \hspace*{-0.6cm} 
  \includegraphics[width=7.4in]{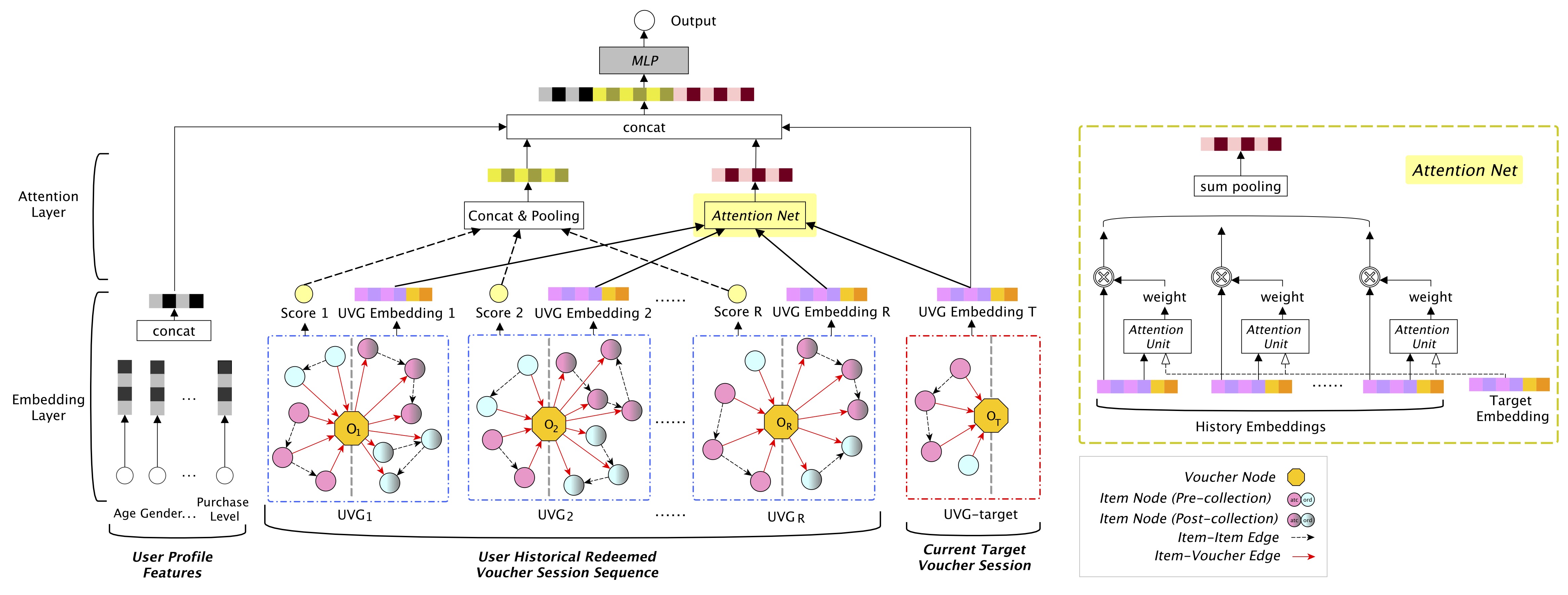}
  \setlength{\abovecaptionskip}{-0.3cm}
  \setlength{\belowcaptionskip}{-0.2cm}
  \caption{The architecture of DMBGN. The input consists of user profile features, UVG sequences representing user historical redeemed voucher sessions, and target voucher session. For each UVG, an embedding and its score are obtained. An attention layer is built on top of the UVG sequences to obtain the aggregated embeddings. The outputs from the three parts are concatenated and forwarded to the MLP for the VRR prediction of the target voucher $O_T$.}
  \label{fig:arch}
\end{figure*}

This section introduces the DMBGN model for VRR prediction. It starts with the problem definition, followed by the illustration of User-Behavior Voucher Graph (UVG) and the overview of the model architecture. Next, details about the graph neural networks on UVG are given. Then, the attention mechanism which aggregates the UVG sequences is explained. Finally, the loss function used for model training is presented.

\subsection{Problem Setup}
Given a set $<\mathcal{U}, \mathcal{O}, \mathcal{I}, \mathcal{C} >$, $\mathcal{U}=\{u_1,u_2,u_3,\cdots,u_m\}$ denotes the set of $m$ users, $\mathcal{O}=\{o_1,o_2,o_3,\cdots,o_n\}$ denotes the set of $n$ vouchers, $\mathcal{I}=\{i_1,i_2,i_3,\cdots,i_t\}$ denotes the set of $t$ items, $\mathcal{C}^{u,o}=\{c^{u,o}_1,c^{u,o}_2,c^{u,o}_3,\cdots,c^{u,o}_k\}$ denotes $k$ types of behavior happening before and after user $u$ collects voucher $o$. A behavior is an item from $\mathcal{I}$, associated with an action type (e.g, click, add-to-cart, order). Further, $\mathcal{C}^{u,o}$ is divided into two subsets $\mathcal{C}^{u,o}_b$ and $\mathcal{C}^{u,o}_a$, representing the behavior happening before ($b$) and after ($a$) voucher collection, respectively. In $\mathcal{C}^{u,o}_a$, all the behaviors after the voucher collection and before the voucher redemption (or expiration) are included. In $\mathcal{C}^{u,o}_b$, the nearest behavior items before voucher collection, up to a specified number, are included. A voucher session describes a voucher collected by the user and the associated behavior, which can be defined as a tuple of $S(u, o, \mathcal{C}^{u,o})$. 

The problem of VRR predication is to calculate the probability of a user to redeem a voucher, given the behaviors before user collects the voucher. Formally, the voucher redemption rate is defined as $P(y_{RR} = 1 | u \in \mathcal{U}, o\in \mathcal{O}, \mathcal{C}^{u,o}_b) $, where $y_{RR}=1$ if and only if user $u$ redeems the voucher $o$.

\subsection{User-Behavior Voucher Graph (UVG)}
\label{UVG construction}
To model the user-voucher-item relationships, a User-Behavior Voucher Graph (UVG) is constructed for each voucher session $S(u, o, \mathcal{C}^{u,o})$. The graph is defined as $\mathcal{G}=(\mathcal{V},\mathcal{E})$ where $\mathcal{V}$ represents the node set and $\mathcal{E}$ represents the edge set. $\mathcal{V}$ contains a node for the voucher, denoted as $\mathcal{V}_o$ and the nodes for items in $\mathcal{C}^{u,o}$, denoted as $\mathcal{V}_I$. Item nodes $\mathcal{V}_I$ can be further distinguished according to action timestamps as the pre-collection ($b$) and post-collection ($a$) phases, and according to the action types, i.e., add-to-cart ($atc$) and order ($ord$) \footnote{We also conducted experiments considering nodes related to $click$ action type, but the results do not show any further improvement. This may attribute to the noises in the large amount of clicked items, and the weak correlations between clicked items and collected vouchers.}. Thus, $\mathcal{V}_I$ is divided into four sub-groups $\mathcal{V}_{I,atc}^b$, $\mathcal{V}_{I,ord}^b$, $\mathcal{V}_{I,atc}^a$ and $\mathcal{V}_{I,ord}^a$ along the two dimensions. 

Edges in $\mathcal{E}$ have two types. The first type connects the item nodes within the above four sub-groups of $\mathcal{V}_I$, while the other type connects the item nodes to the voucher node. As shown in Fig. \ref{fig:sub-graph}, a UVG is a satellite-like structure where the center is the voucher node surrounded by four zones. These four zones represent four types of relations, corresponding to the four sub-groups of $\mathcal{V}_{I}$, respectively. The central voucher node is only connected to the closest (in event timestamp) $Z$ item nodes in every zone, while in each particular zone, an item node is connected to its closest node. UVG is a directed graph, and the direction of each edge follows the corresponding chronological order.

\subsection{Model Architecture}
\label{Framework}
The overall framework of DMBGN is presented in Fig. \ref{fig:arch}. The model takes user’s profile information, historical voucher sessions information, and the current voucher session information as inputs. The left bottom part is for the user profiles features handled by an embedding layer. The central bottom is for the user's historical redeemed voucher sessions, while the right bottom is for the session of the target voucher to be predicted. The information and relationships about the voucher and its surrounding behavior are represented by UVG. Graph Neural Networks (GNN) are applied on the UVG to generate the embedding, as well as a score reflecting the quality of the embedding. On top of a user's historical UVG sequence, an attention network is introduced to obtain the aggregated embedding, which represents the user's voucher usage interest with respect to the target voucher. An addition output of the UVG sequence is also created by concatenating and pooling all the UVG scores in the sequence. 

The outputs of the user profiles, historical voucher sessions and target voucher session are concatenated together and forwarded to an MLP layer for final redemption rate prediction. The result is a 0-to-1 value indicating the probability that a user would redeem a voucher. The parameters of DMBGN are learned on labelled training samples and can be applied for the later VRR prediction.

\subsection{Graph Embedding for UVG}
\label{Graph Embedding for UVG}
\subsubsection{\textbf{Higher-order Graph Neural Networks}}
When each UVG is constructed, graph neural networks are applied to extract the high-level representations. Higher-order Graph Neural Networks with Weisfeiler-Leman Algorithm \cite{morris2019weisfeiler} have the advantage of fast computation, easy tuning and robustness, and have been widely used in production pipelines. Hence Higher-order GNN is adopted here to conduct the convolution operations which update each node $v \in \mathcal{V}$ in UVG as follows:
\begin{equation}\label{equ:conv}
{f^{(l)}}(v) = \sigma \big( {f^{(l - 1)}}(v) \cdot W_1^{(l)} + \sum\limits_{w \in N(v)} {{f^{(l - 1)}}(w) \cdot W_2^{(l)} \big)}
\vspace{-0.1cm}
\end{equation} where $l$ represents $l^{th}$ layer, and $N(v)$ represents the neighbors of $v$. $W_1^{(l)}$ and $W_2^{(l)}$ represent the GNN weight parameters for $l^{th}$ convolution layer. $\sigma(\cdot)$ represents the sigmoid activation function.

The first layer node embedding $f^{(0)}(v)$ is obtained by separate embedding layers for the input item and voucher features. For an item, the input features include item id, category id and item price level; for a voucher, the input features include the corresponding marketing activity id and schema values (i.e., minimum-spend and discount-amount).

The node embeddings generated by Equation (\ref{equ:conv}) are further aggregated into the user-behavior embedding $e_{UVG,b}$, the voucher embedding $e_{UVG,o}$, and the final UVG embedding $e_{UVG}$. Recall the item nodes in UVG can be divided into four sub-groups. We calculate the embedding for a sub-group $g$ as follows:
\begin{equation}\label{equ:pool}
\begin{split}
{e_g} &= {e_{g,avg}~} || {~e_{g,max}}
\end{split}
\end{equation} where $e_{g,avg}$ and $e_{g,max}$ represent the average and the maximization of all the node embeddings in the sub-group $g$. These two results are concatenated to form the sub-group's final embedding $e_g$. Next, the embeddings of four sub-groups, denoted as $atc\_{pre}$, $ord\_{pre}$, $atc\_{post}$ and $ord\_{post}$ are concatenated and forwarded to an MLP to derive $e_{UVG,b}$, that is:
\begin{equation}\label{equ:pool UVG}
\begin{split}
e_{UVG, b} &= \textit{MLP}(atc\_{pre} ~||~ ord\_{pre} ~||~ atc\_{post} ~||~ ord\_{post})
\end{split}
\end{equation}

The voucher embedding $e_{UVG,o}$ is simply the embedding from the voucher node. $e_{UVG,b}$ and $e_{UVG,o}$ are concatenated to obtain the final UVG embedding $e_{UVG}$ as in Equation (\ref{equ:eUVG}). In addition, a score for a historical UVG is generated from the dot-product between $e_{UVG,b}$ and $e_{UVG,o}$ as in Equation (\ref{equ:sUVG}). The score, namely $s_{UVG}$, represents how well the voucher embedding alone can reflect the related behavior associated with this voucher.
\begin{equation}\label{equ:eUVG}
\begin{split}
e_{UVG} &= e_{UVG,b}~ || ~e_{UVG, o}
\end{split}
\end{equation}

\vspace{-0.6cm}
\begin{equation}\label{equ:sUVG}
\begin{split}
s_{UVG} &= \sigma(e_{UVG,b} \cdot e_{UVG, o})
\end{split}
\end{equation} where the $\sigma(\cdot)$ represents the sigmoid function.

The first-layer node embeddings $f^{(0)}(v)$ for items and vouchers can either be learned from scratch with the main task, or be trained independently in advance. In this work, we pre-train these initial item node embeddings and voucher embeddings separately.

\subsubsection{\textbf{Pre-training of Item Embedding}}
\label{SGNS item emb}
Following the previous works in \cite{pfadler2020billion,vasile2016meta,perozzi2014deepwalk}, a meta-path type graph embedding algorithm is applied to generate the item embeddings. In order to avoid overfitting, the algorithm is run on all the logged user behavior data from our platform. To enrich the expression of users' item sequences, side-information features such as $category\_id$, $brand\_id$, and $shop\_id$ are added for training. The pre-training is conducted separately for add-to-cart and order behavior in this paper, so that the same item of different behavior types has different embeddings. The learned embeddings are used to initialize the item nodes in both historical and target UVGs.

\subsubsection{\textbf{Pre-training of Voucher Embedding}}
The historical UVGs in DMBGN represent the voucher-redeemed sessions. However, there may exist an overwhelming number of non-redeemed voucher sessions in the voucher usage data set. In this work, we utilize all the redeemed and non-redeemed voucher sessions from the platform to pre-train the voucher embeddings. A learned voucher embedding is expected to be capable of representing its related usage behavior, so it should be close to its corresponding behavior embedding in a semantic space if the voucher is redeemed, and far away otherwise. 

For each user voucher session, a UVG is constructed as described in Section \ref{UVG construction}. The item embeddings input to UVG are implemented as in Section \ref{SGNS item emb}, while the voucher embedding is randomly initialized. Recall that $s_{UVG}$, the score of the UVG represents the similarity between $e_{UVG,b}$ and $e_{UVG,o}$. Loss function for voucher embedding training is defined as:
\vspace{-0.2cm}
\begin{equation}
L_{pre-voucher} = \frac{1}{N} \sum \limits_{n = 1}^N l(y_{UVG}^n, s_{UVG}^n)
\end{equation} where $N$ represents the total size of the UVG pre-trained samples, and $n$ represents the $n^{th}$ sample. $y_{UVG}$ is a 1-or-0 label of UVG indicating the ground-truth whether the user redeems this voucher or not. $l(\cdot)$ represents the binary cross-entropy loss function. 

After training, the initial embeddings of the voucher nodes are learned. Note that the above pre-training involves the ground-truth label whether a user redeems a voucher. Therefore, it is only conducted in the historical UVGs, but not in the target UVG. For all the target UVGs, the voucher embeddings are randomly initialized and learned from scratch with the main task.

In addition to the voucher embeddings, the weight parameters of the Higher-order Graph Neural Networks are learned during the pre-training of UVGs. These weights are used for DMBGN network initialization and to be further fine-tuned with the main task.

\subsection{Attention Networks on UVG sequences}
A user may have multiple previous voucher redemption sessions and thus may have a sequence of historical UVGs. To discover the user's long-term voucher usage preference (e.g. ``What kind of voucher the user will be interested in?'' and ``What kinds of items will the user purchases with the voucher?''), an attention layer is added on top of the historical and the target UVGs. The \textit{AttentionNet} proposed in the work of DIN \cite{zhou2018deep} is adopted as follows:
\begin{equation}\label{equ:att}
\begin{split}
    w^r &= AttentionUnit(e_{UVG}^r,e_{UVG}^T,\phi ), r = 1,2,\cdots, R \\
    {h_T} &= \sum\nolimits_{r = 1}^R {w^r \cdot {e_{UVG}^r}}\\
\end{split}
\end{equation} here, $e_{UVG}^r$ and $e_{UVG}^T$ are the embeddings of the $r^{th}$ historical UVG and the target UVG, respectively, and $h_T$ is the high-level representation of historical UVG sequences with respect to the target voucher $O_T$. The \textit{AttentionUnit} in Fig. \ref{fig:arch} is a feed-forward neural network with scalar output $w^r$ as the attention weight, and the network parameters are represented as $\phi$. Apart from the two input embedding vectors, \textit{AttentionUnit} adds their outer-product and difference as well, which helps model the interactions between an input pair as mentioned in the original DIN model. 

The learned attention network gives higher weights to the historical voucher sessions that are semantically closer to the target session. These historical sessions include the complete behavior information, and thus could better reflect the user’s long-term voucher usage preference with respect to the target voucher. Referring to such sessions gives valuable clues whether a voucher will be redeemed in the end.

\subsection{Loss Function}
To train the DMBGN model, two loss functions are considered. The first one, $L_{target}$, evaluates the loss based on the difference between the predicted voucher redemption result $\hat y_{RR}^n$ and the ground-truth label $y_{RR}^n$, where $n$ represents the $n^{th}$ training sample. The loss function is formulated as:
\begin{equation}
{L_{target}} = \frac{1}{N}\sum\limits_{n = 1}^N { - (y_{RR}^n} \log \hat y_{RR}^n + (1 - y_{RR}^n)\log (1 - \hat y_{RR}^n))
% L_{target} = l({{\hat y}}_{RR},y_{RR}).
\end{equation} where $N$ is the total number of training samples.

Another loss function is introduced for regularization. Recall in DMBGN, each historical UVG session outputs a score $s_{UVG}$ indicating whether the voucher and the behavior embeddings are semantically close. It is expected that the voucher embedding can reflect the related behavior, the score should be as close to 1 as possible because each historical UVG corresponds to a redeemed session. Inspired by the work in \cite{zhou2019deep}, for better supervision of the UVG graph network training during the main task, the auxiliary loss function is defined as:
\begin{equation}
L_{aux} = \frac{1}{N}\sum\limits_{n = 1}^N \big(\frac{1}{R_n} \sum\limits_{r = 1}^{R_n} {-log(s_{UVG}^r}) \big)
\end{equation} where $R_n$ denotes the history UVG sequence length from the $n^{th}$ sample, and $r$ represents the $r^{th}$ UVG in a user's history UVG sequence. Therefore, the final loss $L$ is the weighted sum of $L_{target}$ and $L_{aux}$, represented as follows:
\begin{equation}
L = L_{target}+\alpha \cdot L_{aux}
\end{equation} where $\alpha$ is the trade-off coefficient.

\section{Experiments}
\subsection{Dataset}
For the VRR prediction task, there is no public dataset so far. In this work, we use the large-scale dataset from the real voucher log data from Lazada, a leading South-East Asia (SEA) E-commerce platform of Alibaba Group. The data covers Lazada’s voucher marketing activities during campaign periods such as Double 11 and Double 12 from 2019 to 2020. We select the data from three countries in SEA, named as Region-A, Region-B and Region-C, respectively. The three datasets contain users' platform voucher collection and redemption logs as well as users' behavior occurring both before and after voucher collection. To preserve adequate amount of user behavior without introducing too much noise, the 95\% percentile of users' behavior length is retained. For each user, at most 45 add-to-cart and 20 order items before or after voucher collection are reserved. The three datasets are of different sizes, and reflect various user-voucher engagement degrees. Overall, 44.2\% (Region-A), 11.6\% (Region-B) and 31.2\% (Region-C) of the samples have at least one previous voucher redemption record. 

Each sample in the dataset is represented as a unique <$user\_id$, $voucher\_id$> pair, with a label 1 or 0 indicating if the user redeemed the voucher or not. A $voucher\_id$ represents an individual voucher marketing activity. Each sample is associated with the user profile features (predicted age level, predicted gender, predicted purchase level, and user's order statistics), voucher features (minimum-spend, discount-amount, collection and redemption timestamps) and item features (pre-trained embeddings, brand\_id, category\_id, price level). All numerical features used in the following experiments are normalized through z-normalization and all sparse id features are encoded using LabelEncoder.

We have published a desensitized subset of original dataset Region-C for public use. Table \ref{tab:dataset} summarizes the statistics of three datasets, a full description of the dataset is also available on GitHub.

\begin{table}
  \hspace*{-0.3cm} 
  \scalebox{1}{
  \caption{Statistics of three datasets from different regions.}
  \label{tab:dataset}
  \footnotesize
  \begin{tabular}{ccccccc}
    \toprule
    \multirow{2}{*}{Dataset} & \multirow{2}{*}{Sessions} & \multirow{2}{*}{Users} & \multirow{2}{*}{Items} & \multirow{2}{*}{Vouchers} & Training & Testing \\
    &&&&&Samples&Samples\\
    \midrule
    Region-A & 3.4M & 730K & 7.2M & 1k & 2.7M & 687K\\
    Region-B & 14M & 6.5M& 19M & 0.5k & 11M & 2.8M\\
    Region-C & 24M & 7.0M & 15.5M & 1k & 19.4M & 4.9M\\
    \bottomrule
\end{tabular}
}
\vspace{-0.3cm}
\end{table}

\subsection{Compared Models}
\label{compared models}
To evaluate the performance of proposed method, we compare DMBGN with the following models which are commonly adopted in item CTR prediction tasks.

\begin{itemize}
    \item {\textbf{LR:}}
    Logistic Regression \cite{LR_ref} is a shallow model. Only dense features from user profiles and voucher information are used in the model.
    \item{\textbf{GBDT:}}
    Gradient Boosting Decision Tree \cite{ke2017lightgbm} is used to assess the performance of non deep-learning algorithms. Similarly, only dense features are used.
    \item {\textbf{DNN:}} The Deep Neural Network is used as the first baseline taking both dense features and embedding of sparse id features into the model.
    \item {\textbf{WDL:}} Wide and Deep model \cite{cheng2016wide} is widely accepted in real industrial applications. Compared with DNN, it has an additional linear model besides the deep model. All dense and sparse features are used in both wide and deep side.
    \item {\textbf{DIN:}}
    Deep Interest Network \cite{zhou2018deep} is an attention-based model in recommendation systems that has been proven successful in Alibaba. We use this as our second baseline, replacing the user's historical item sequences with user's historical voucher sequences to adapt to the VRR prediction task. All dense and sparse features are used in the model along with the attention part.
\end{itemize}

\vspace{-0.1cm}
\begin{table*}[h]
  \setlength\abovecaptionskip{-5pt}
  \setlength\belowcaptionskip{-5pt}
  \caption{Model performance evaluated by AUC on three production datasets. The Relative Improvement (RelaImpr) takes DNN and DIN as baseline models. Best results of all methods are highlighted in boldface.}
  \label{tab:main}
  \resizebox{1.03\linewidth}{!}{
  \hspace*{-0.6cm}
  \begin{tabular}{ccccccccccccccc}
    \toprule
    \multirow{3}{*}{Model} & \multicolumn{4}{c}{Region-A} &  & \multicolumn{4}{c}{Region-B} &  & \multicolumn{4}{c}{Region-C}\\
    \cline{2-5}\cline{7-10}\cline{12-15}
    & AUC & RelaImpr & RelaImpr & Logloss & & AUC & RelaImpr & RelaImpr & Logloss & & AUC & RelaImpr & RelaImpr & Logloss\\
    & mean $\pm$ std & (DNN) & (DIN) & mean & & mean $\pm$ std & (DNN) & (DIN) & mean & & mean $\pm$ std & (DNN) & (DIN) & mean \\
    \midrule
     LR & 0.7491 $\pm$ 0.00005 & -9.04\% & -14.43\% & 0.4385 & & 0.7805 $\pm$ 0.00009 & -10.74\% & -17.93\% & 0.3712 & & 0.7565 $\pm$ 0.00011 & -16.66\% & -26.39\% & 0.3202\\
     GBDT & 0.7538 $\pm$ 0.00004 & -7.34\% & -12.83\% & 0.4215 & & 0.7924 $\pm$ 0.00004 & -6.94\% & -14.44\% & 0.3610 & & 0.7982 $\pm$ 0.00005 & -3.12\% & -14.43\% & 0.2933\\
     DNN & 0.7739 $\pm$ 0.00000  & 0.00\% & -5.93\% & 0.4156 & & 0.8142 $\pm$ 0.00000 & 0.00\% & -8.06\% & 0.3478 & & 0.8078 $\pm$ 0.00000 & 0.00\% & -11.67\% & 0.2966\\
     WDL & 0.7752 $\pm$ 0.00004 & 0.47\% & -5.49\% & 0.4145 & & 0.8188 $\pm$ 0.00004 & 1.46\% & -6.72\% & 0.3452 & & 0.8110 $\pm$ 0.00009 & 1.05\% & -10.74\% & 0.2948\\
     DIN & 0.7912 $\pm$ 0.00029 & 6.30\% & 0.00\% & 0.4021 & & 0.8417 $\pm$ 0.00017 & 8.77\% & 0.00\% & 0.3248 & & 0.8485 $\pm$ 0.00022 & 13.22\% & 0.00\% & 0.2692\\
     DMBGN-AvgPooling & 0.7961 $\pm$ 0.00017 & 8.09\% & 1.68\% & 0.3975 & & 0.8430 $\pm$ 0.00015 & 9.15\% & 0.36\% & 0.3238 & & 0.8497 $\pm$ 0.00026 & 13.60\% & 0.34\% & 0.2681\\
     DMBGN-Pretrained & 0.8012 $\pm$ 0.00014 & 9.97\% & 3.45\%  & 0.3954 & & 0.8444 $\pm$ 0.00011 & 9.62\% & 0.78\% & 0.3217 & & 0.8530 $\pm$ 0.00023 & 14.67\% & 1.29\% & 0.2658\\
     \textbf{DMBGN} & \textbf{0.8034 $\pm$ 0.00023} & \textbf{10.76\%} & \textbf{4.20\%} & \textbf{0.3921} & & \textbf{0.8486 $\pm$ 0.00021} & \textbf{10.94\%} & \textbf{2.00\%} & \textbf{0.3184} & & \textbf{0.8591 $\pm$ 0.00015} & \textbf{16.65\%} & \textbf{3.04\%} & \textbf{0.2614}\\
  \bottomrule
\end{tabular}
}
\vspace{-0.3cm}
\end{table*}

\subsection{Evaluation Metrics}
Two widely used metrics for binary classification tasks are adopted in our experiments: AUC and Logloss. Area under receiver operator curve (AUC) measures the probability of ranking a random positive sample higher than a negative one. Logloss is also utilized which reflects the prediction inaccuracy of voucher redemption rate. Each experiment is repeated 5 times. Mean and standard deviation of AUC and mean of Logloss are reported.

In addition, to measure the relative AUC improvement over models, RelaImpr \cite{zhou2018deep, yan2014coupled} is also used which is defined as: 
\begin{equation}\label{equ:relaimpr}
\setlength{\belowdisplayskip}{3pt}
{RelaImpr} = \left(\frac{AUC(measured~model)-0.5}{AUC(base~model)-0.5}-1 \right) \times 100\%
\end{equation}

\subsection{Hyper-parameter Settings}
All experiments in this paper use Adam optimizer with $\beta_1=0.9$, $\beta_2=0.999$ and learning rate $lr = 0.001$. 8 GPUs are used in the experiment with batch size of 1024. In the \textit{MLP} layer across different deep models, the hidden units are (128, 64); dropout rate is set as 0.5; $L2$ regularization is applied to avoid over-fitting with coefficient as 0.1. In \textit{Attention Net}, PReLU \cite{zhou2018deep} is selected as the activation function with one layer of 64 hidden units. The embedding dimension of sparse id features is 16. The Graph Neural Networks for UVGs adopt a graph convolution layer and a top-K pooling layer \cite{gao2019graph} with ratio = 0.9. The loss trade-off coefficient is set as $\alpha = 1.0$.

Besides, as 99.10\% of users have not more than 6 historical redeemed vouchers, the maximal length of historical UVG sequence is set as $R=6$. The number of nodes connected to the voucher node in UVG mentioned in Section \ref{UVG construction} is set as $Z=6$.

\section{Results and Discussions}
In this section, we analyze the overall performance of DMBGN and other compared models. The following four questions are also proposed to be answered:

\begin{itemize}
    \item \textbf{Q1:} Are graph neural networks in DMBGN helpful in improving the overall performance?
    \item \textbf{Q2:} How does the user behavior happening before and after voucher collection help improve the overall performance of DMBGN?
    \item \textbf{Q3:} How does user activeness in terms of historical UVG sequence length affect the DMBGN performance? 
    \item \textbf{Q4:} How are the user-voucher-item structural relationships distilled by Higher-order Graph Neural Networks?
\end{itemize}

The experimental results of DMBGN and the other compared models are summarized in Table \ref{tab:main}. Besides, we also list the results of two variants of DMBGN, which use the same network structure but different embedding generation methods:
\begin{itemize}
    \item \textbf{AvgPooling}: Instead of using Higher-order Graph Neural Networks to model user-voucher-item relationships, it directly takes an average of pre-trained item embeddings from user behavior happening both before and after voucher collection. For target UVG, it only takes an average of pre-collection item embeddings.
    \item \textbf{Pretrained}: It uses the same weight parameters of Higher-order GNN learned during the voucher embedding pre-training as mentioned in Section \ref{Framework}. But the values of weight parameters are not further updated during the main task training. 
    \item \textbf{DMBGN}: The proposed model in this work, which loads the pre-trained GNN network including the item and voucher node embeddings. The GNN network parameters are further fine-tuned according to the final training loss.
\end{itemize}
\vspace{-0.35cm}

\subsection{Overall Performance}
Based on the results in Table \ref{tab:main}, we have the following observations:
\begin{itemize}
    \item DMBGN consistently outperforms other baseline models across all three datasets in terms of AUC and Logloss, the standard deviation of AUC is smaller than 0.001 indicating reliable results. Especially, DMBGN achieves over 10$\%$ to 16$\%$ relative improvement in AUC compared to the most commonly used DNN model. It supports that DMBGN is capable of capturing the complex user-voucher-item relationship with User-Behavior Voucher Graph (UVG) well, and such relationship is commonly ignored in other models.
    \item Compared with the competitive model DIN, DMBGN achieves around 2\% to 4\% relative uplift in AUC. Recall that in the VRR tasks, DMBGN does not simply replace items by vouchers, but constructs the UVG and utilizes graph neural networks to better model the pattern between user behaviors and voucher collection.
    \item We observe that the AUC of DMBGN improves from 0.8034 to 0.8591 as the training size grows from 3.4 million to 24 million, indicating that DMBGN is capable of handling large-scale dataset, and performs better with more training samples involved. Across three datasets, the AUC relative improvement against DIN is larger in regions that are more active in voucher usage, specifically, 4.20\% uplift in Region-A, compared to 2.00\% in Region-B and 3.04\% in Region-C.
\end{itemize}

\subsection{Ablation Study (Q1)}
To further demonstrate the effectiveness by introducing the graph structure, we compare the performance of DMBGN with its two variants. The results are shown in the last three rows in Table \ref{tab:main}:
\begin{itemize}
    \item Taking \textit{AvgPooling} as baseline, \textit{Pre-trained} achieves the relative AUC improvement of 1.7\%, 0.4\%, 0.9\% as calculated in Equation (\ref{equ:relaimpr}) on dataset Region-A, B, C respectively. Recall that \textit{AvgPooling} and \textit{Pre-trained} take different strategies to generate the user behavior embedding $e_{UVG,b}$, by simple mean pooling and by graph techniques, respectively. The out-performance of \textit{Pre-trained} over \textit{AvgPooling} supports that the construction of $UVG$ and Higher-order GNN could better capture user-voucher-item relationships.
    \item After fine-tuning of Higher-order GNN network parameters, DMBGN further improves the AUC over \textit{Pre-trained} version by 0.7\%, 1.2\%, 1.7\% on dataset Region-A, B, C, respectively. This improvement indicates that the graph network can adaptively learn from the final loss.
\end{itemize}

\subsection{Post-collection User Behavior Study (Q2)}
Unlike item-wise recommendation, the user behavior related to voucher collection plays an important role in final redemption decisions. Hence, we conduct experiment to compare the influence from behavior in different phases (i.e., pre-collection and post-collection) on the overall performance of DMBGN. The results are summarized in Table \ref{tab:after collection}.

\vspace{-0.15cm}
\begin{table}[h]
    \setlength{\abovecaptionskip}{-1cm}
    \Huge
    \setlength{\belowcaptionskip}{-0.4cm}
    \caption{Experiment results of DMBGN for including and excluding post-collection user behavior on three datasets.}
    \label{tab:after collection}
    \resizebox{\linewidth}{!}{
        \begin{tabular}{cccccccccccc}
            \toprule
            \multirow{2}{*}{Model}&\multicolumn{2}{c}{Region-A}&&\multicolumn{2}{c}{Region-B}&&\multicolumn{2}{c}{Region-C}\\
            \cline{2-3}\cline{5-6}\cline{8-9}
            & AUC & Logloss & & AUC & Logloss & & AUC & Logloss\\
            \midrule
            DMBGN w/ pre-collection & \multirow{2}{*}{0.8014} & \multirow{2}{*}{0.3937} & & \multirow{2}{*}{0.8479} & \multirow{2}{*}{0.3191} & & \multirow{2}{*}{0.8557}&\multirow{2}{*}{0.2639} \\behavior only&&&&&&\\
            \hline
            DMBGN w/ pre- and post- & \multirow{2}{*}{0.8034} & \multirow{2}{*}{0.3921} & & \multirow{2}{*}{0.8486} & \multirow{2}{*}{0.3184} & & \multirow{2}{*}{0.8591} &\multirow{2}{*}{0.2614}\\ collection behavior&&&&&&\\
            \bottomrule
        \end{tabular}
    }
    \vspace{-0.3cm}
\end{table}
\vspace{-0.15cm}

Based on the results shown in Table \ref{tab:after collection}, we see that DBMGN using pre-collection behavior alone could achieve relative AUC improvement of 3.5\%, 1.8\%, 2.1\% across three datasets compared to DIN, which uses $voucher\_id$ but without related user behavior. Moreover, adding the post-collection behavior could further improve the overall performance by 0.7\%, 0.2\%, 0.9\% compared to using the pre-collection only. Those pre-collection user behavior usually reflects a user's  specific interest during a voucher campaign, and the post-collection could further reflect the interest stimulated by the current collected voucher. The results indicate that DBMGN is able to capture user's interest in both before and after voucher collection for better VRR prediction.

\subsection{Analysis of User Activeness (Q3)}
\begin{figure}[h]
  \centering
  \vspace{-0.2cm}
  \setlength{\abovecaptionskip}{-0.005cm}
  \setlength{\belowcaptionskip}{-0.1cm}
  \includegraphics[width=2in]{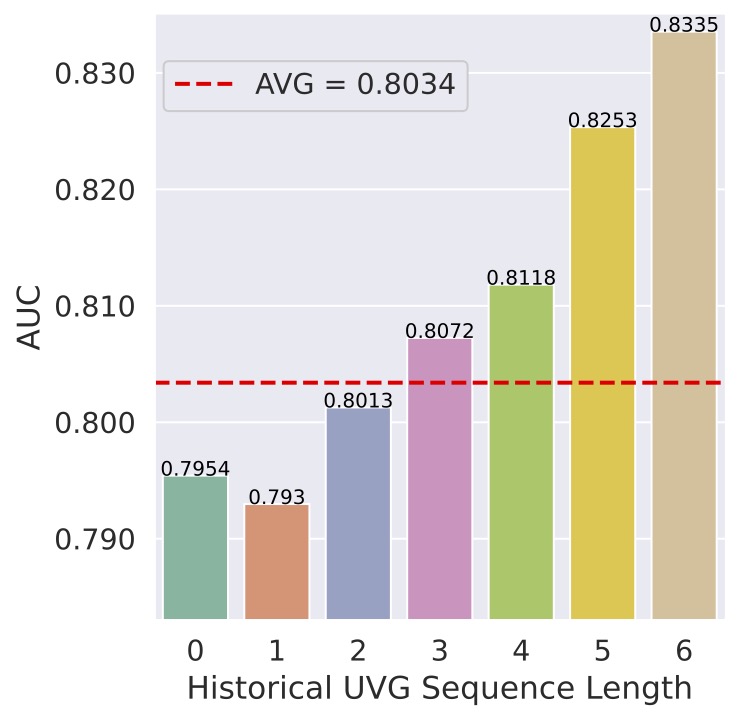}
  \caption{AUC performance on different historical UVG sequence length.}
  \label{fig:sequence length}
  \vspace{-0.1cm}
\end{figure}

Experiments are performed to see how DMBGN is affected by users' activeness in terms of the historical UVG sequence length. Region-A which is the most active region is taken for analysis. Fig. \ref{fig:sequence length} visualizes the test AUC with respect to various historical UVG sequence lengths from 0 to 6 (the max historical UVG sequence length in our system setting). Note that the test data size under each sequence length group is more than 10k, which is statistically adequate for this analysis.

In Fig. \ref{fig:sequence length}, as the historical UVG sequence length increases, the corresponding AUC increases from 0.7930 to 0.8335. This observation indicates that the attention structure in DMBGN can effectively capture users' long-term interest in voucher usage. For the groups with historical sequence length larger than 1, where the attention mechanism actually matters, the performance is better than those with smaller historical length. It is interesting to observe that the AUC of the group with length 1 is slightly lower than that of the group with sequence 0. A possible explanation is that the users from the former group are also inactive in voucher usage, and only one single redeemed session might not be sufficient to reflect the user's voucher preference, but nevertheless introduces noise.

\subsection{\mbox{User-behavior Embedding Visualization (Q4)}}
\begin{center}
    \begin{figure}[h]
        \includegraphics[width=\columnwidth, keepaspectratio,trim={0 0 3.2cm 0},clip]{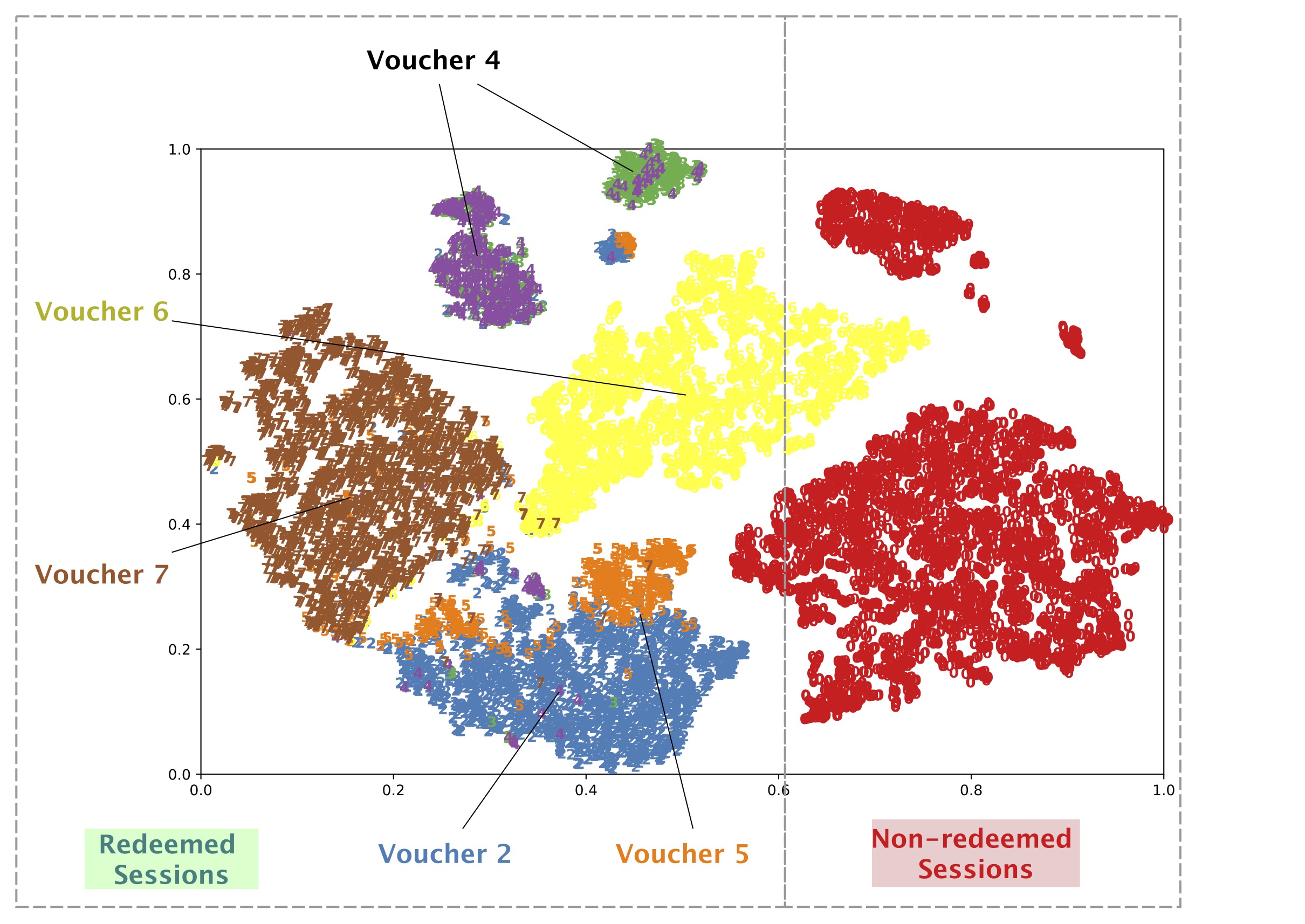}
        \caption{Visualization of learned user-behavior embedding vectors $e_{UVG,b}$ from UVG. Red color represents non-redeemed sessions, and the rest colors represent redeemed sessions with respect to different voucher ids.}
        \label{fig:visualization}
        \vspace{-0.4cm}
    \end{figure}
\end{center}

To show the effectiveness of knowledge distilling by the Higher-order Graph Neural Networks, we randomly sample the sessions of five $voucher\_ids$ from dataset Region-B, and visualize the corresponding learned user-behavior embedding vectors $e_{UVG,b}$ from UVG in 2 dimension using t-SNE \cite{vanDerMaaten2008} algorithm in Fig. \ref{fig:visualization}. All non-redeemed sessions from those $voucher\_ids$ are marked in red color, while the redeemed sessions with the same $voucher\_id$ are marked with an individual color. It shows that the learned embeddings can well discriminate redeemed and non-redeemed sessions. It also illustrates that sessions with the same color are mostly located in the same cluster. One exception is Voucher 4 cluster, which is divided into two sub-clusters. A further analysis shows that the split clusters correspond to users of different activeness levels. Specifically, in the purple sub-cluster, 93.3\% users have greater than or equal to 19 total items related to user behavior while in the green sub-cluster, 83.1\% users have less than 19 items. Overall, Fig. \ref{fig:visualization} indicates that the learned embeddings from DMBGN can well represent the users' voucher-related behavior.

\vspace{-0.3cm}
\vspace{-0.5cm}

\section{Conclusion and Future Work}
In this paper, we propose a voucher redemption prediction (VRR) model: Deep Multi-Behavior Graph Networks (DMBGN). It considers the historical user behavior happening both before and after voucher collection to make better inference. To capture the user-voucher-item structural relationships, a User-Behavior Voucher Graph (UVG) is constructed, and Higher-order Graph Neural Networks are applied to extract the structural features reflecting interactions between voucher and user behavior. An attention network is built on top of the UVG sequences to mine user's long-term voucher usage preference. Extensive experiments in three large-scale datasets are conducted. The results show that DMBGN achieves 10\% to 16\% relative uplift in AUC compared to DNN, as well as 2\% to 4\% relative uplift in AUC compared to DIN. This demonstrates that DMBGN is effective for the VRR prediction task.

There are several potential future works for exploration. In a voucher campaign, there may exist multiple vouchers to be distributed simultaneously or within a short time interval. The influence of different vouchers on user behavior are intertwined. More sophisticated graph structures are needed to capture voucher-voucher cooperative or conflicting relationships. Besides, the applicable condition of a voucher may have restrictions over specific shops, brands or categories. The influence of the voucher usage scopes on the user behavior is also worthwhile to be explored. 

\bibliographystyle{ACM-Reference-Format}
\bibliography{reference}

%%% -*-BibTeX-*-
%%% Do NOT edit. File created by BibTeX with style
%%% ACM-Reference-Format-Journals [18-Jan-2012].

\begin{thebibliography}{38}

%%% ====================================================================
%%% NOTE TO THE USER: you can override these defaults by providing
%%% customized versions of any of these macros before the \bibliography
%%% command.  Each of them MUST provide its own final punctuation,
%%% except for \shownote{}, \showDOI{}, and \showURL{}.  The latter two
%%% do not use final punctuation, in order to avoid confusing it with
%%% the Web address.
%%%
%%% To suppress output of a particular field, define its macro to expand
%%% to an empty string, or better, \unskip, like this:
%%%
%%% \newcommand{\showDOI}[1]{\unskip}   % LaTeX syntax
%%%
%%% \def \showDOI #1{\unskip}           % plain TeX syntax
%%%
%%% ====================================================================

\ifx \showCODEN    \undefined \def \showCODEN     #1{\unskip}     \fi
\ifx \showDOI      \undefined \def \showDOI       #1{#1}\fi
\ifx \showISBNx    \undefined \def \showISBNx     #1{\unskip}     \fi
\ifx \showISBNxiii \undefined \def \showISBNxiii  #1{\unskip}     \fi
\ifx \showISSN     \undefined \def \showISSN      #1{\unskip}     \fi
\ifx \showLCCN     \undefined \def \showLCCN      #1{\unskip}     \fi
\ifx \shownote     \undefined \def \shownote      #1{#1}          \fi
\ifx \showarticletitle \undefined \def \showarticletitle #1{#1}   \fi
\ifx \showURL      \undefined \def \showURL       {\relax}        \fi
% The following commands are used for tagged output and should be
% invisible to TeX
\providecommand\bibfield[2]{#2}
\providecommand\bibinfo[2]{#2}
\providecommand\natexlab[1]{#1}
\providecommand\showeprint[2][]{arXiv:#2}

\bibitem[\protect\citeauthoryear{Agrawal and Devanur}{Agrawal and
  Devanur}{2015}]%
        {agrawal2015linear}
\bibfield{author}{\bibinfo{person}{Shipra Agrawal} {and}
  \bibinfo{person}{Nikhil~R Devanur}.} \bibinfo{year}{2015}\natexlab{}.
\newblock \showarticletitle{Linear contextual bandits with knapsacks}.
\newblock \bibinfo{journal}{\emph{arXiv preprint arXiv:1507.06738}}
  (\bibinfo{year}{2015}).
\newblock


\bibitem[\protect\citeauthoryear{Bao, Wen, Li, Liu, Lin, and Yang}{Bao
  et~al\mbox{.}}{2020}]%
        {bao2020gmcm}
\bibfield{author}{\bibinfo{person}{Wentian Bao}, \bibinfo{person}{Hong Wen},
  \bibinfo{person}{Sha Li}, \bibinfo{person}{Xiao-Yang Liu},
  \bibinfo{person}{Quan Lin}, {and} \bibinfo{person}{Keping Yang}.}
  \bibinfo{year}{2020}\natexlab{}.
\newblock \showarticletitle{Gmcm: Graph-based micro-behavior conversion model
  for post-click conversion rate estimation}. In
  \bibinfo{booktitle}{\emph{Proceedings of the 43rd International ACM SIGIR
  Conference on Research and Development in Information Retrieval}}.
  \bibinfo{pages}{2201--2210}.
\newblock


\bibitem[\protect\citeauthoryear{Chakrabarty, Zhou, and Lukose}{Chakrabarty
  et~al\mbox{.}}{2008}]%
        {chakrabarty2008online}
\bibfield{author}{\bibinfo{person}{Deeparnab Chakrabarty},
  \bibinfo{person}{Yunhong Zhou}, {and} \bibinfo{person}{Rajan Lukose}.}
  \bibinfo{year}{2008}\natexlab{}.
\newblock \showarticletitle{Online knapsack problems}. In
  \bibinfo{booktitle}{\emph{Workshop on internet and network economics
  (WINE)}}.
\newblock


\bibitem[\protect\citeauthoryear{Chen, Zhang, Zhang, Ma, Liu, and Ma}{Chen
  et~al\mbox{.}}{2020}]%
        {Chen2020EfficientHC}
\bibfield{author}{\bibinfo{person}{C. Chen}, \bibinfo{person}{Min Zhang},
  \bibinfo{person}{Yongfeng Zhang}, \bibinfo{person}{W. Ma},
  \bibinfo{person}{Yiqun Liu}, {and} \bibinfo{person}{Shaoping Ma}.}
  \bibinfo{year}{2020}\natexlab{}.
\newblock \showarticletitle{Efficient Heterogeneous Collaborative Filtering
  without Negative Sampling for Recommendation}. In
  \bibinfo{booktitle}{\emph{AAAI}}.
\newblock


\bibitem[\protect\citeauthoryear{Chen, Zhao, Li, et~al\mbox{.}}{Chen
  et~al\mbox{.}}{2019}]%
        {chen2019behavior}
\bibfield{author}{\bibinfo{person}{Qiwei Chen}, \bibinfo{person}{Huan Zhao},
  \bibinfo{person}{Wei Li}, {et~al\mbox{.}}} \bibinfo{year}{2019}\natexlab{}.
\newblock \showarticletitle{Behavior sequence transformer for e-commerce
  recommendation in alibaba}. In \bibinfo{booktitle}{\emph{Proceedings of the
  1st International Workshop on Deep Learning Practice for High-Dimensional
  Sparse Data}}. \bibinfo{pages}{1--4}.
\newblock


\bibitem[\protect\citeauthoryear{Cheng, Koc, Harmsen, Shaked, Chandra, Aradhye,
  Anderson, Corrado, Chai, Ispir, et~al\mbox{.}}{Cheng et~al\mbox{.}}{2016}]%
        {cheng2016wide}
\bibfield{author}{\bibinfo{person}{Heng-Tze Cheng}, \bibinfo{person}{Levent
  Koc}, \bibinfo{person}{Jeremiah Harmsen}, \bibinfo{person}{Tal Shaked},
  \bibinfo{person}{Tushar Chandra}, \bibinfo{person}{Hrishi Aradhye},
  \bibinfo{person}{Glen Anderson}, \bibinfo{person}{Greg Corrado},
  \bibinfo{person}{Wei Chai}, \bibinfo{person}{Mustafa Ispir}, {et~al\mbox{.}}}
  \bibinfo{year}{2016}\natexlab{}.
\newblock \showarticletitle{Wide \& deep learning for recommender systems}. In
  \bibinfo{booktitle}{\emph{Proceedings of the 1st workshop on deep learning
  for recommender systems}}. \bibinfo{pages}{7--10}.
\newblock


\bibitem[\protect\citeauthoryear{Dong, Chawla, and Swami}{Dong
  et~al\mbox{.}}{2017}]%
        {dong2017metapath2vec}
\bibfield{author}{\bibinfo{person}{Yuxiao Dong}, \bibinfo{person}{Nitesh~V
  Chawla}, {and} \bibinfo{person}{Ananthram Swami}.}
  \bibinfo{year}{2017}\natexlab{}.
\newblock \showarticletitle{metapath2vec: Scalable representation learning for
  heterogeneous networks}. In \bibinfo{booktitle}{\emph{Proceedings of the 23rd
  ACM SIGKDD international conference on knowledge discovery and data mining}}.
  \bibinfo{pages}{135--144}.
\newblock


\bibitem[\protect\citeauthoryear{Feng, Hu, Lv, Liu, Zhang, and Ou}{Feng
  et~al\mbox{.}}{2020}]%
        {feng2020atbrg}
\bibfield{author}{\bibinfo{person}{Yufei Feng}, \bibinfo{person}{Binbin Hu},
  \bibinfo{person}{Fuyu Lv}, \bibinfo{person}{Qingwen Liu},
  \bibinfo{person}{Zhiqiang Zhang}, {and} \bibinfo{person}{Wenwu Ou}.}
  \bibinfo{year}{2020}\natexlab{}.
\newblock \showarticletitle{ATBRG: Adaptive Target-Behavior Relational Graph
  Network for Effective Recommendation}. In
  \bibinfo{booktitle}{\emph{Proceedings of the 43rd International ACM SIGIR
  Conference on Research and Development in Information Retrieval}}.
  \bibinfo{pages}{2231--2240}.
\newblock


\bibitem[\protect\citeauthoryear{Gai and Qiu}{Gai and Qiu}{2018}]%
        {gai2018optimal}
\bibfield{author}{\bibinfo{person}{Keke Gai} {and} \bibinfo{person}{Meikang
  Qiu}.} \bibinfo{year}{2018}\natexlab{}.
\newblock \showarticletitle{Optimal resource allocation using reinforcement
  learning for IoT content-centric services}.
\newblock \bibinfo{journal}{\emph{Applied Soft Computing}}
  \bibinfo{volume}{70} (\bibinfo{year}{2018}), \bibinfo{pages}{12--21}.
\newblock


\bibitem[\protect\citeauthoryear{Gao, He, Gan, Chen, Feng, Li, Chua, and
  Jin}{Gao et~al\mbox{.}}{2019}]%
        {8731537}
\bibfield{author}{\bibinfo{person}{Chen Gao}, \bibinfo{person}{Xiangnan He},
  \bibinfo{person}{Dahua Gan}, \bibinfo{person}{Xiangning Chen},
  \bibinfo{person}{Fuli Feng}, \bibinfo{person}{Yong Li},
  \bibinfo{person}{Tat-Seng Chua}, {and} \bibinfo{person}{Depeng Jin}.}
  \bibinfo{year}{2019}\natexlab{}.
\newblock \showarticletitle{Neural Multi-task Recommendation from
  Multi-behavior Data}. In \bibinfo{booktitle}{\emph{2019 IEEE 35th
  International Conference on Data Engineering (ICDE)}}.
  \bibinfo{pages}{1554--1557}.
\newblock
\urldef\tempurl%
\url{https://doi.org/10.1109/ICDE.2019.00140}
\showDOI{\tempurl}


\bibitem[\protect\citeauthoryear{Gao and Ji}{Gao and Ji}{2019}]%
        {gao2019graph}
\bibfield{author}{\bibinfo{person}{Hongyang Gao} {and}
  \bibinfo{person}{Shuiwang Ji}.} \bibinfo{year}{2019}\natexlab{}.
\newblock \showarticletitle{Graph u-nets}. In
  \bibinfo{booktitle}{\emph{international conference on machine learning}}.
  PMLR, \bibinfo{pages}{2083--2092}.
\newblock


\bibitem[\protect\citeauthoryear{Grover and Leskovec}{Grover and
  Leskovec}{2016}]%
        {grover2016node2vec}
\bibfield{author}{\bibinfo{person}{Aditya Grover} {and} \bibinfo{person}{Jure
  Leskovec}.} \bibinfo{year}{2016}\natexlab{}.
\newblock \showarticletitle{node2vec: Scalable feature learning for networks}.
  In \bibinfo{booktitle}{\emph{Proceedings of the 22nd ACM SIGKDD international
  conference on Knowledge discovery and data mining}}.
  \bibinfo{pages}{855--864}.
\newblock


\bibitem[\protect\citeauthoryear{Guo, Tang, Ye, Li, and He}{Guo
  et~al\mbox{.}}{2017}]%
        {DeepFM}
\bibfield{author}{\bibinfo{person}{Huifeng Guo}, \bibinfo{person}{Ruiming
  Tang}, \bibinfo{person}{Yunming Ye}, \bibinfo{person}{Zhenguo Li}, {and}
  \bibinfo{person}{Xiuqiang He}.} \bibinfo{year}{2017}\natexlab{}.
\newblock \showarticletitle{DeepFM: A Factorization-Machine Based Neural
  Network for CTR Prediction}. In \bibinfo{booktitle}{\emph{Proceedings of the
  26th International Joint Conference on Artificial Intelligence}} (Melbourne,
  Australia) \emph{(\bibinfo{series}{IJCAI'17})}. \bibinfo{publisher}{AAAI
  Press}, \bibinfo{pages}{1725–1731}.
\newblock
\showISBNx{9780999241103}


\bibitem[\protect\citeauthoryear{Hamilton, Ying, and Leskovec}{Hamilton
  et~al\mbox{.}}{2017}]%
        {hamilton2017inductive}
\bibfield{author}{\bibinfo{person}{William~L Hamilton}, \bibinfo{person}{Rex
  Ying}, {and} \bibinfo{person}{Jure Leskovec}.}
  \bibinfo{year}{2017}\natexlab{}.
\newblock \showarticletitle{Inductive representation learning on large graphs}.
\newblock \bibinfo{journal}{\emph{arXiv preprint arXiv:1706.02216}}
  (\bibinfo{year}{2017}).
\newblock


\bibitem[\protect\citeauthoryear{Hu, Tang, Chen, et~al\mbox{.}}{Hu
  et~al\mbox{.}}{2018}]%
        {hu2018promotion}
\bibfield{author}{\bibinfo{person}{Wan-Hsun Hu}, \bibinfo{person}{Shih-Hsien
  Tang}, \bibinfo{person}{Yin-Che Chen}, {et~al\mbox{.}}}
  \bibinfo{year}{2018}\natexlab{}.
\newblock \showarticletitle{Promotion recommendation method and system based on
  random forest}. In \bibinfo{booktitle}{\emph{Proceedings of the 5th
  Multidisciplinary International Social Networks Conference}}.
  \bibinfo{pages}{1--5}.
\newblock


\bibitem[\protect\citeauthoryear{Jin, Gao, He, Jin, and Li}{Jin
  et~al\mbox{.}}{2020}]%
        {jin2020multi}
\bibfield{author}{\bibinfo{person}{Bowen Jin}, \bibinfo{person}{Chen Gao},
  \bibinfo{person}{Xiangnan He}, \bibinfo{person}{Depeng Jin}, {and}
  \bibinfo{person}{Yong Li}.} \bibinfo{year}{2020}\natexlab{}.
\newblock \showarticletitle{Multi-behavior recommendation with graph
  convolutional networks}. In \bibinfo{booktitle}{\emph{Proceedings of the 43rd
  International ACM SIGIR Conference on Research and Development in Information
  Retrieval}}. \bibinfo{pages}{659--668}.
\newblock


\bibitem[\protect\citeauthoryear{Ke, Meng, Finley, Wang, Chen, Ma,
  et~al\mbox{.}}{Ke et~al\mbox{.}}{2017}]%
        {ke2017lightgbm}
\bibfield{author}{\bibinfo{person}{Guolin Ke}, \bibinfo{person}{Qi Meng},
  \bibinfo{person}{Thomas Finley}, \bibinfo{person}{Taifeng Wang},
  \bibinfo{person}{Wei Chen}, \bibinfo{person}{Weidong Ma}, {et~al\mbox{.}}}
  \bibinfo{year}{2017}\natexlab{}.
\newblock \showarticletitle{Lightgbm: A highly efficient gradient boosting
  decision tree}.
\newblock \bibinfo{journal}{\emph{Advances in neural information processing
  systems}}  \bibinfo{volume}{30} (\bibinfo{year}{2017}),
  \bibinfo{pages}{3146--3154}.
\newblock


\bibitem[\protect\citeauthoryear{Kellerer, Pferschy, and Pisinger}{Kellerer
  et~al\mbox{.}}{2004}]%
        {kellerer2004multiple}
\bibfield{author}{\bibinfo{person}{Hans Kellerer}, \bibinfo{person}{Ulrich
  Pferschy}, {and} \bibinfo{person}{David Pisinger}.}
  \bibinfo{year}{2004}\natexlab{}.
\newblock \showarticletitle{The multiple-choice knapsack problem}.
\newblock In \bibinfo{booktitle}{\emph{Knapsack Problems}}.
  \bibinfo{publisher}{Springer}, \bibinfo{pages}{317--347}.
\newblock


\bibitem[\protect\citeauthoryear{Kipf and Welling}{Kipf and Welling}{2016}]%
        {kipf2016semi}
\bibfield{author}{\bibinfo{person}{Thomas~N Kipf} {and} \bibinfo{person}{Max
  Welling}.} \bibinfo{year}{2016}\natexlab{}.
\newblock \showarticletitle{Semi-supervised classification with graph
  convolutional networks}.
\newblock \bibinfo{journal}{\emph{arXiv preprint arXiv:1609.02907}}
  (\bibinfo{year}{2016}).
\newblock


\bibitem[\protect\citeauthoryear{Li, Sun, Weng, Huo, and Ren}{Li
  et~al\mbox{.}}{2020}]%
        {li2020spending}
\bibfield{author}{\bibinfo{person}{Liangwei Li}, \bibinfo{person}{Liucheng
  Sun}, \bibinfo{person}{Chenwei Weng}, \bibinfo{person}{Chengfu Huo}, {and}
  \bibinfo{person}{Weijun Ren}.} \bibinfo{year}{2020}\natexlab{}.
\newblock \showarticletitle{Spending Money Wisely: Online Electronic Coupon
  Allocation based on Real-Time User Intent Detection}. In
  \bibinfo{booktitle}{\emph{Proceedings of the 29th ACM International
  Conference on Information \& Knowledge Management}}.
  \bibinfo{pages}{2597--2604}.
\newblock


\bibitem[\protect\citeauthoryear{Li, Tarlow, Brockschmidt, and Zemel}{Li
  et~al\mbox{.}}{2015}]%
        {li2015gated}
\bibfield{author}{\bibinfo{person}{Yujia Li}, \bibinfo{person}{Daniel Tarlow},
  \bibinfo{person}{Marc Brockschmidt}, {and} \bibinfo{person}{Richard Zemel}.}
  \bibinfo{year}{2015}\natexlab{}.
\newblock \showarticletitle{Gated graph sequence neural networks}.
\newblock \bibinfo{journal}{\emph{arXiv preprint arXiv:1511.05493}}
  (\bibinfo{year}{2015}).
\newblock


\bibitem[\protect\citeauthoryear{Ma, Zhao, Yi, Chen, Hong, and Chi}{Ma
  et~al\mbox{.}}{2018}]%
        {ma2018modeling}
\bibfield{author}{\bibinfo{person}{Jiaqi Ma}, \bibinfo{person}{Zhe Zhao},
  \bibinfo{person}{Xinyang Yi}, \bibinfo{person}{Jilin Chen},
  \bibinfo{person}{Lichan Hong}, {and} \bibinfo{person}{Ed~H Chi}.}
  \bibinfo{year}{2018}\natexlab{}.
\newblock \showarticletitle{Modeling task relationships in multi-task learning
  with multi-gate mixture-of-experts}. In \bibinfo{booktitle}{\emph{Proceedings
  of the 24th ACM SIGKDD International Conference on Knowledge Discovery \&
  Data Mining}}. \bibinfo{pages}{1930--1939}.
\newblock


\bibitem[\protect\citeauthoryear{McMahan, Holt, Sculley, Young, Ebner,
  et~al\mbox{.}}{McMahan et~al\mbox{.}}{2013}]%
        {LR_ref}
\bibfield{author}{\bibinfo{person}{H.~Brendan McMahan}, \bibinfo{person}{Gary
  Holt}, \bibinfo{person}{D. Sculley}, \bibinfo{person}{Michael Young},
  \bibinfo{person}{Dietmar Ebner}, {et~al\mbox{.}}}
  \bibinfo{year}{2013}\natexlab{}.
\newblock \showarticletitle{Ad Click Prediction: A View from the Trenches}. In
  \bibinfo{booktitle}{\emph{Proceedings of the 19th ACM SIGKDD International
  Conference on Knowledge Discovery and Data Mining}} (Chicago, Illinois, USA)
  \emph{(\bibinfo{series}{KDD '13})}. \bibinfo{publisher}{Association for
  Computing Machinery}, \bibinfo{address}{New York, NY, USA},
  \bibinfo{pages}{1222–1230}.
\newblock
\showISBNx{9781450321747}
\urldef\tempurl%
\url{https://doi.org/10.1145/2487575.2488200}
\showDOI{\tempurl}


\bibitem[\protect\citeauthoryear{Morris, Ritzert, Fey, Hamilton, Lenssen,
  Rattan, and Grohe}{Morris et~al\mbox{.}}{2019}]%
        {morris2019weisfeiler}
\bibfield{author}{\bibinfo{person}{Christopher Morris}, \bibinfo{person}{Martin
  Ritzert}, \bibinfo{person}{Matthias Fey}, \bibinfo{person}{William~L
  Hamilton}, \bibinfo{person}{Jan~Eric Lenssen}, \bibinfo{person}{Gaurav
  Rattan}, {and} \bibinfo{person}{Martin Grohe}.}
  \bibinfo{year}{2019}\natexlab{}.
\newblock \showarticletitle{Weisfeiler and leman go neural: Higher-order graph
  neural networks}. In \bibinfo{booktitle}{\emph{Proceedings of the AAAI
  Conference on Artificial Intelligence}}, Vol.~\bibinfo{volume}{33}.
  \bibinfo{pages}{4602--4609}.
\newblock


\bibitem[\protect\citeauthoryear{Niu, Li, Li, Xiao, Sun, Deng, and Chen}{Niu
  et~al\mbox{.}}{2020}]%
        {niu2020dual}
\bibfield{author}{\bibinfo{person}{Xichuan Niu}, \bibinfo{person}{Bofang Li},
  \bibinfo{person}{Chenliang Li}, \bibinfo{person}{Rong Xiao},
  \bibinfo{person}{Haochuan Sun}, \bibinfo{person}{Hongbo Deng}, {and}
  \bibinfo{person}{Zhenzhong Chen}.} \bibinfo{year}{2020}\natexlab{}.
\newblock \showarticletitle{A Dual Heterogeneous Graph Attention Network to
  Improve Long-Tail Performance for Shop Search in E-Commerce}. In
  \bibinfo{booktitle}{\emph{Proceedings of the 26th ACM SIGKDD International
  Conference on Knowledge Discovery \& Data Mining}}.
  \bibinfo{pages}{3405--3415}.
\newblock


\bibitem[\protect\citeauthoryear{Perozzi, Al-Rfou, and Skiena}{Perozzi
  et~al\mbox{.}}{2014}]%
        {perozzi2014deepwalk}
\bibfield{author}{\bibinfo{person}{Bryan Perozzi}, \bibinfo{person}{Rami
  Al-Rfou}, {and} \bibinfo{person}{Steven Skiena}.}
  \bibinfo{year}{2014}\natexlab{}.
\newblock \showarticletitle{Deepwalk: Online learning of social
  representations}. In \bibinfo{booktitle}{\emph{Proceedings of the 20th ACM
  SIGKDD international conference on Knowledge discovery and data mining}}.
  \bibinfo{pages}{701--710}.
\newblock


\bibitem[\protect\citeauthoryear{Pfadler, Zhao, Wang, Wang, Huang, and
  Lee}{Pfadler et~al\mbox{.}}{2020}]%
        {pfadler2020billion}
\bibfield{author}{\bibinfo{person}{Andreas Pfadler}, \bibinfo{person}{Huan
  Zhao}, \bibinfo{person}{Jizhe Wang}, \bibinfo{person}{Lifeng Wang},
  \bibinfo{person}{Pipei Huang}, {and} \bibinfo{person}{Dik~Lun Lee}.}
  \bibinfo{year}{2020}\natexlab{}.
\newblock \showarticletitle{Billion-scale Recommendation with Heterogeneous
  Side Information at Taobao}. In \bibinfo{booktitle}{\emph{2020 IEEE 36th
  International Conference on Data Engineering (ICDE)}}. IEEE,
  \bibinfo{pages}{1667--1676}.
\newblock


\bibitem[\protect\citeauthoryear{Qiongyu}{Qiongyu}{2020}]%
        {qiongyu2020prediction}
\bibfield{author}{\bibinfo{person}{Shi Qiongyu}.}
  \bibinfo{year}{2020}\natexlab{}.
\newblock \showarticletitle{Prediction of O2O Coupon Usage Based on XGBoost
  Model}. In \bibinfo{booktitle}{\emph{2020 The 11th International Conference
  on E-business, Management and Economics}}. \bibinfo{pages}{33--36}.
\newblock


\bibitem[\protect\citeauthoryear{van~der Maaten and Hinton}{van~der Maaten and
  Hinton}{2008}]%
        {vanDerMaaten2008}
\bibfield{author}{\bibinfo{person}{Laurens van~der Maaten} {and}
  \bibinfo{person}{Geoffrey Hinton}.} \bibinfo{year}{2008}\natexlab{}.
\newblock \showarticletitle{Visualizing Data using {t-SNE}}.
\newblock \bibinfo{journal}{\emph{Journal of Machine Learning Research}}
  \bibinfo{volume}{9} (\bibinfo{year}{2008}), \bibinfo{pages}{2579--2605}.
\newblock
\urldef\tempurl%
\url{http://www.jmlr.org/papers/v9/vandermaaten08a.html}
\showURL{%
\tempurl}


\bibitem[\protect\citeauthoryear{Vasile, Smirnova, and Conneau}{Vasile
  et~al\mbox{.}}{2016}]%
        {vasile2016meta}
\bibfield{author}{\bibinfo{person}{Flavian Vasile}, \bibinfo{person}{Elena
  Smirnova}, {and} \bibinfo{person}{Alexis Conneau}.}
  \bibinfo{year}{2016}\natexlab{}.
\newblock \showarticletitle{Meta-prod2vec: Product embeddings using
  side-information for recommendation}. In
  \bibinfo{booktitle}{\emph{Proceedings of the 10th ACM Conference on
  Recommender Systems}}. \bibinfo{pages}{225--232}.
\newblock


\bibitem[\protect\citeauthoryear{Veličković, Cucurull, Casanova, Romero,
  et~al\mbox{.}}{Veličković et~al\mbox{.}}{2018}]%
        {velivckovic2017graph}
\bibfield{author}{\bibinfo{person}{Petar Veličković},
  \bibinfo{person}{Guillem Cucurull}, \bibinfo{person}{Arantxa Casanova},
  \bibinfo{person}{Adriana Romero}, {et~al\mbox{.}}}
  \bibinfo{year}{2018}\natexlab{}.
\newblock \bibinfo{title}{Graph Attention Networks}.
\newblock
\newblock
\showeprint[arxiv]{1710.10903}~[stat.ML]


\bibitem[\protect\citeauthoryear{Wang, Fu, Fu, and Wang}{Wang
  et~al\mbox{.}}{2017}]%
        {wang2017deep}
\bibfield{author}{\bibinfo{person}{Ruoxi Wang}, \bibinfo{person}{Bin Fu},
  \bibinfo{person}{Gang Fu}, {and} \bibinfo{person}{Mingliang Wang}.}
  \bibinfo{year}{2017}\natexlab{}.
\newblock \showarticletitle{Deep \& cross network for ad click predictions}.
\newblock In \bibinfo{booktitle}{\emph{Proceedings of the ADKDD'17}}.
  \bibinfo{pages}{1--7}.
\newblock


\bibitem[\protect\citeauthoryear{Wen, Zhang, Wang, Lv, Bao, Lin, and Yang}{Wen
  et~al\mbox{.}}{2020}]%
        {wen2020entire}
\bibfield{author}{\bibinfo{person}{Hong Wen}, \bibinfo{person}{Jing Zhang},
  \bibinfo{person}{Yuan Wang}, \bibinfo{person}{Fuyu Lv},
  \bibinfo{person}{Wentian Bao}, \bibinfo{person}{Quan Lin}, {and}
  \bibinfo{person}{Keping Yang}.} \bibinfo{year}{2020}\natexlab{}.
\newblock \showarticletitle{Entire space multi-task modeling via post-click
  behavior decomposition for conversion rate prediction}. In
  \bibinfo{booktitle}{\emph{Proceedings of the 43rd International ACM SIGIR
  Conference on Research and Development in Information Retrieval}}.
  \bibinfo{pages}{2377--2386}.
\newblock


\bibitem[\protect\citeauthoryear{Xia, Huang, Xu, Dai, Zhang, et~al\mbox{.}}{Xia
  et~al\mbox{.}}{2020}]%
        {xia2020multiplex}
\bibfield{author}{\bibinfo{person}{Lianghao Xia}, \bibinfo{person}{Chao Huang},
  \bibinfo{person}{Yong Xu}, \bibinfo{person}{Peng Dai}, \bibinfo{person}{Bo
  Zhang}, {et~al\mbox{.}}} \bibinfo{year}{2020}\natexlab{}.
\newblock \showarticletitle{Multiplex Behavioral Relation Learning for
  Recommendation via Memory Augmented Transformer Network}. In
  \bibinfo{booktitle}{\emph{Proceedings of the 43rd International ACM SIGIR
  Conference on Research and Development in Information Retrieval}}.
  \bibinfo{pages}{2397--2406}.
\newblock


\bibitem[\protect\citeauthoryear{Xu, He, Tan, Li, Lang, and Guo}{Xu
  et~al\mbox{.}}{2020}]%
        {xu2020deep}
\bibfield{author}{\bibinfo{person}{Weinan Xu}, \bibinfo{person}{Hengxu He},
  \bibinfo{person}{Minshi Tan}, \bibinfo{person}{Yunming Li},
  \bibinfo{person}{Jun Lang}, {and} \bibinfo{person}{Dongbai Guo}.}
  \bibinfo{year}{2020}\natexlab{}.
\newblock \showarticletitle{Deep Interest with Hierarchical Attention Network
  for Click-Through Rate Prediction}. In \bibinfo{booktitle}{\emph{Proceedings
  of the 43rd International ACM SIGIR Conference on Research and Development in
  Information Retrieval}}. \bibinfo{pages}{1905--1908}.
\newblock


\bibitem[\protect\citeauthoryear{Yan, Li, Xue, and Han}{Yan
  et~al\mbox{.}}{2014}]%
        {yan2014coupled}
\bibfield{author}{\bibinfo{person}{Ling Yan}, \bibinfo{person}{Wu-jun Li},
  \bibinfo{person}{Gui-Rong Xue}, {and} \bibinfo{person}{Dingyi Han}.}
  \bibinfo{year}{2014}\natexlab{}.
\newblock \showarticletitle{Coupled group lasso for web-scale ctr prediction in
  display advertising}. In \bibinfo{booktitle}{\emph{International Conference
  on Machine Learning}}. PMLR, \bibinfo{pages}{802--810}.
\newblock


\bibitem[\protect\citeauthoryear{Zhou, Mou, Fan, Pi, Bian, Zhou,
  et~al\mbox{.}}{Zhou et~al\mbox{.}}{2019}]%
        {zhou2019deep}
\bibfield{author}{\bibinfo{person}{Guorui Zhou}, \bibinfo{person}{Na Mou},
  \bibinfo{person}{Ying Fan}, \bibinfo{person}{Qi Pi}, \bibinfo{person}{Weijie
  Bian}, \bibinfo{person}{Chang Zhou}, {et~al\mbox{.}}}
  \bibinfo{year}{2019}\natexlab{}.
\newblock \showarticletitle{Deep interest evolution network for click-through
  rate prediction}. In \bibinfo{booktitle}{\emph{Proceedings of the AAAI
  conference on artificial intelligence}}, Vol.~\bibinfo{volume}{33}.
  \bibinfo{pages}{5941--5948}.
\newblock


\bibitem[\protect\citeauthoryear{Zhou, Zhu, Song, Fan, Zhu, Ma, Yan, Jin, Li,
  and Gai}{Zhou et~al\mbox{.}}{2018}]%
        {zhou2018deep}
\bibfield{author}{\bibinfo{person}{Guorui Zhou}, \bibinfo{person}{Xiaoqiang
  Zhu}, \bibinfo{person}{Chenru Song}, \bibinfo{person}{Ying Fan},
  \bibinfo{person}{Han Zhu}, \bibinfo{person}{Xiao Ma},
  \bibinfo{person}{Yanghui Yan}, \bibinfo{person}{Junqi Jin},
  \bibinfo{person}{Han Li}, {and} \bibinfo{person}{Kun Gai}.}
  \bibinfo{year}{2018}\natexlab{}.
\newblock \showarticletitle{Deep interest network for click-through rate
  prediction}. In \bibinfo{booktitle}{\emph{Proceedings of the 24th ACM SIGKDD
  International Conference on Knowledge Discovery \& Data Mining}}.
  \bibinfo{pages}{1059--1068}.
\newblock


\end{thebibliography}

\end{document}